\documentclass[aps,pra,twocolumn,notitlepage,superscriptaddress]{revtex4-1}
\usepackage[breaklinks=true]{hyperref}

\usepackage{graphicx,color}
\usepackage{dcolumn}
\usepackage{bm}
\usepackage{amsmath,amssymb,mathrsfs}
\usepackage{url}
\usepackage{hyperref}

\usepackage{xcolor}

\newcommand{\R}{\mathbb{R}}
\newcommand{\C}{\mathbb{C}}

\newcommand{\grp}[1]{\mathsf{#1}}
\newcommand{\spc}[1]{\mathcal{#1}}

\def\d{{\rm d}}

\newcommand{\Span}{{\mathsf{Span}}}
\newcommand{\Lin}{\mathsf{Lin}}
\newcommand{\Herm}{\mathsf{Herm}}

\def\>{\rangle}
\def\<{\langle}

\newcommand{\map}[1]{\mathcal{#1}}
\newcommand{\Tr}{\operatorname{Tr}}

\newcommand{\St}{{\mathsf{St}}}

\newcommand{\hS}{{\widehat {\map S} }}

\newtheorem{theo}{Theorem}

\newtheorem{lemma}{Lemma}
\newtheorem{prop}{Proposition}

\def\Proof{{\bf Proof.~}}
\def\qed{$\blacksquare$ \newline}

\usepackage{qcircuit}

\begin{document}
    \title{Virtual quantum broadcasting}
    \author{Arthur J.\ Parzygnat}
    \email{arthurjp@mit.edu}
    \thanks{AJP and JF contributed equally to this work.}
    \affiliation{Department of Mathematics, Massachusetts Institute of Technology, Cambridge, Massachusetts 02139, USA} 
    \affiliation{Experimental Study Group, Massachusetts Institute of Technology, Cambridge, Massachusetts 02139, USA}
    \author{James Fullwood}
    \email{fullwood@sjtu.edu.cn}
    \affiliation{School of Mathematical Sciences, Shanghai Jiao Tong University, 800 Dongchuan Road, Shanghai, People's Republic of China}
    \author{Francesco Buscemi}
    \email{buscemi@i.nagoya-u.ac.jp}
    \affiliation{Graduate School of Informatics, Nagoya University, Chikusa-ku, 464-8601 Nagoya, Japan}
    \author{Giulio Chiribella}
    \email{giulio@cs.hku.hk}
    \affiliation{QICI Quantum Information and Computation Initiative, Department of Computer Science, The University of Hong Kong, Pok Fu Lam Road, Hong Kong}
    \affiliation{Department of Computer Science, University of Oxford, Wolfson Building, Parks Road, Oxford, UK}
    \affiliation{Perimeter Institute for Theoretical Physics, 31 Caroline Street North, Waterloo,  Ontario, Canada}
  
\begin{abstract}
The  quantum no-broadcasting theorem states that it is impossible to produce  perfect copies of an arbitrary quantum state, even if the copies are allowed to be correlated.   Here we show that, although quantum broadcasting cannot be achieved by any physical process, it can be achieved by a virtual process, described by a Hermitian-preserving  trace-preserving map.  This virtual process is canonical: it is the only map that broadcasts all quantum states, is covariant under unitary evolution, is invariant under permutations of the copies,   and reduces to the classical broadcasting map when subjected to decoherence. We show that the optimal physical approximation to the canonical broadcasting map is the optimal universal quantum cloning, and we also show that virtual broadcasting can be achieved by a virtual measure-and-prepare protocol, where a virtual measurement is performed, and, depending on the outcomes, two copies of a virtual quantum state are generated. Finally, we use canonical virtual broadcasting to prove a uniqueness result for quantum states over time.    
       \end{abstract}

  \maketitle

{\em Introduction.~}In standard quantum theory, physical processes are described by completely positive trace-preserving (CPTP) linear maps~\cite{Kr83}. The no-broadcasting theorem says that no physical process can broadcast an unknown quantum state to two parties so that each party would obtain the same statistics, via local measurements, as would be obtained from the original state~\cite{BCFJS96,KaHe08,Ma10,PaBayes}. In fact, the theorem also applies to positive maps, without the requirement of complete positivity~\cite{BBLW07}. The situation changes, however, if one lifts the positivity requirement. 

In this Letter, we study broadcasting maps from the larger set of {\em Hermitian-preserving trace-preserving (HPTP) maps}, that is, linear maps that transform Hermitian operators into Hermitian operators while preserving the trace.    HPTP maps have applications in the context of error mitigation~\cite{TBG17,JWW21,TEMG22}, simulating non-Markovian dynamics~\cite{RMAD23}, two-point correlation functions~\cite{BDOV13,BDOV14}, and quantum states over time~\cite{FuPa22,FuPa22a,LiNg23},  while their Hilbert--Schmidt adjoints have been studied in the context of retrodicting quantum observables~\cite{Pe71,Ts22,Ts22b}. Physically, HPTP maps can be simulated by sampling over a set of quantum processes and then suitably post-processing the measurement statistics on the corresponding output states~\cite{BDOV13,BDOV14,RMAD23}. 

The main result of this Letter is Theorem~\ref{theo:unique}, which states that a unique HPTP quantum broadcasting map is singled out by three natural  requirements: covariance under unitary evolution, invariance under permutation of the output systems, and consistency with the unique classical cloning channel. We explicitly construct this canonical quantum broadcasting map, and we also establish two distinct physical interpretations for it. 

First, Theorem~\ref{theo:closest-physical} states that the optimal  universal symmetric  cloning process in quantum theory can be interpreted as the quantum channel that best approximates the canonical broadcasting map. 
  Second, Theorem~\ref{theo:convex} provides a realization of the canonical broadcasting map as a probabilistic mixture of the completely depolarizing channel and a particular Hermitian operator-valued measure-and-prepare protocol distributed over all pure quantum states of a given system. 

Finally, we show that our uniqueness result for virtual broadcasting yields a uniqueness result for \emph{quantum states over time}~\cite{HHPBS17,FuPa22}. Quantum states over time provide a nascent approach to quantum dynamics that is in close analogy with spacetime physics. They have provided insights to time-reversal symmetry~\cite{FuPa22a}, quantum Bayesian inference~\cite{PaBayes,PaRuBayes,PaQPL21,GPRR21,FuPa22a,PaBu22}, and dynamical measures of quantum information~\cite{FuPa23}. Such quantum states over time also specialize to the pseudo-density matrices associated with timelike-separated Pauli measurements on systems of qubits~\cite{HHPBS17, ZPTGVF18, LQDV23, fullwood2023}. From the perspective of quantum states over time, the output of a virtual broadcasting map is a state over time associated with a quantum state that has evolved according to trivial dynamics. As such, while the no-broadcasting theorem states that quantum broadcasting cannot be realized as a quantum state at a single time, our \emph{virtual} broadcasting theorem implies that quantum broadcasting may be dynamically realized as a quantum state \emph{over} time, and moreover, that such a dynamical realization is unique.     
 
\medskip
{\em Virtual broadcasting maps.}~Let us start by introducing some notation that will be used throughout the paper. For an arbitrary quantum system $S$, we denote by $\spc H_S$ the corresponding Hilbert space, and by $\Lin (S)$ the algebra of all linear operators on $\spc H_S$. The real vector subspace of Hermitian operators will be denoted by $\Herm (S)$, while its affine subspace of unit trace operators will be denoted by $\Herm_1  (S)$.  The convex subspace of all quantum states (positive matrices with unit trace) of system $S$ will be denoted by $\St (S)$. We denote the identity operator in $\Lin (S)$ by $I$ and the the identity transformation acting on $\Lin (S)$ by $\map I$. 

Let $S$, $S_1$, and $S_2$  be three quantum systems all with the same underlying Hilbert space, whose dimension is denoted by $d$.  Let $S_1S_2$ be the composite system made of subsystems $S_1$ and $S_2$, and let $\spc H_{S_1S_2}  = \spc H_{S_1} \otimes \spc H_{S_2}$ be the corresponding Hilbert space.    For an operator $O  \in  \Lin  (S_1S_2)$, the  partial trace over the Hilbert spaces $\spc H_{S_1}$ and $\spc H_{S_2}$ will be denoted by $\Tr_{S_1}[O]$  and $\Tr_{S_2}[O]$, respectively.  A  {\em broadcasting map} from $S$ to  $S_1S_2$ is a linear map $\map B:  \Lin (S)  \to \Lin  (S_1S_2)$ satisfying the conditions 
\begin{align}\label{broadcasting}
  \Tr_{S_1}[ \map B  (\rho)]   =   \Tr_{S_2}[ \map B  (\rho)]   =    \rho  \qquad \forall \rho  \in \Lin (S)\, . 
\end{align} 
We refer to Eq. (\ref{broadcasting}) as the {\em broadcasting condition}. Note that the broadcasting condition  automatically implies that the map 
 $\map B$ must be trace-preserving. 

A Hermitian-preserving map that satisfies the broadcasting condition (\ref{broadcasting}) will be referred to as a {\em virtual broadcasting map}. 
Examples of virtual broadcasting maps abound (see Appendix~\ref{APX0} of the Supplementary Material), and this non-uniqueness is in stark contrast with the properties of broadcasting in classical probability theory, where the assumption of positivity singles out a unique broadcasting map that perfectly copies all pure states of a classical system.  
   
A classical system with $d$ distinct pure states can be thought of as a $d$-dimensional quantum system that has undergone a complete decoherence process with respect to a fixed orthonormal basis  $ \{|i\>\}_{i=1}^{d}$ representing the classical pure states of the system. The {\em decoherence map} with respect to such a basis is then given by the channel $\map D$ defined by sending $\rho$ to $\map D (\rho)  : = \sum_{i=1}^d  \,  \<i|  \rho  | i\> \,  |i\>\<i|$. Given orthonormal bases $\{|i\>_{\rm in}\}_{i=1}^{d_{\rm in}}$ and $\{|j\>_{\rm out}\}_{j=1}^{d_{\rm out}}$ for an input system $S_{\rm in}$ and an output system $S_{\rm out}$, respectively, a {\em classical map} from $S_{\rm in}$ to $S_{\rm out}$ with respect to these bases is a linear map $\map C:  \Lin  (S_{\rm in}) \to \Lin (S_{\rm out})$ satisfying $\map D_{\rm out}\circ \hspace{0.5mm} \map C\circ \hspace{0.5mm} \map D_{\rm in}=\map C$, where $\map D_{\rm in}$ and $\map D_{\rm out}$ are the decoherence maps associated with the given bases for $S_{\rm in}$ and $S_{\rm out}$, respectively. In the broadcasting case, the output system $S_{\rm out}  =  S_1S_2$ consists of two copies of system $S$,  equipped with the product basis  $ \{|i\> \otimes |j\>\}_{i,j=1}^{d}$. A linear map $\map B:\Lin  (S) \to \Lin (S_1S_2)$ is {\em classical} whenever it satisfies $\map B  = (\map D \otimes \map D) \circ \map B \circ \map D$.  A {\em classical broadcasting map} is then a classical map  $\map B_{\rm cl}$  satisfying the broadcasting condition  for the density matrices that are diagonal in the classical  basis, i.e., 
\begin{align}\label{classicalbroadcasting}
 \Tr_{S_1}[ \map B_{\rm cl}  (\rho)]   =     \Tr_{S_2}[ \map B_{\rm cl}  (\rho)]   =    \rho  \quad \forall \rho  \in {\sf Diag} (S)\, \,
\end{align}
where ${\sf Diag} (S) =  \{   \sum_i  \,  p_i \,  |i\>\<i|  ~|~   p_i   \ge 0  \;\forall i\,  ,  \sum_i  p_i=  1  \}$ is the set of diagonal density matrices. The classical broadcasting condition (\ref{classicalbroadcasting}) is satisfied by the map  $\map B_{\rm cl}$ uniquely determined by the condition   
  \begin{align}
 \map B_{\rm cl}   (|i\>\<j|)    = \delta_{ij} |i\>\<i| \otimes |i\>\<i| \qquad \forall  i,j\in  \{1,\dots,  d\} \, ,
\end{align} 
which copies the diagonal pure states $|i\>\<i|$ and their convex combinations, in analogy with the classical copy map, and decoheres off-diagonal elements. 

With the notation in place, we can now state the main result of this Letter, which is that a unique virtual broadcasting map is singled out by three natural requirements. 

The first requirement is 
 {\em covariance under unitary evolution}, which corresponds to the condition   
 \begin{align}\label{covariance}
 \map B (U\rho  U^\dag)   =   (U  \otimes U)  \map B(\rho)  (U\otimes U)^\dag  \qquad \forall \rho, \forall U
 \end{align}
  where  $U\in \Lin (S)$ is an arbitrary unitary operator and  
  $\rho\in\Lin(S)$ is an arbitrary operator. 
  
 The second requirement is   {\em invariance under permutation of the copies}, which corresponds to the condition  
\begin{align}\label{pinvariance}{\tt SWAP}\,  \map B (\rho)\,  {\tt SWAP}  =  \map B(\rho)  \, \qquad   \forall  \rho  \in \Lin (S) \,,
\end{align} 
where  ${\tt SWAP}  \in  \Lin  (S_1S_2)$ is the unitary operator  defined by the relation ${\tt SWAP}   ( |\phi\>  \otimes |\psi\>)  =  |\psi\>\otimes |\phi\> , \forall  |\phi\> ,|\psi\>  \in \C^d $. 

Finally, the third requirement is {\em consistency with classical broadcasting}, which says that if the inputs and outputs of a broadcasting map are subject to decoherence, then the action of the map $\map B$ should coincide with the action of the classical broadcasting map, i.e.,  
\begin{align}\label{restrictstoclassical}
(\map D  \otimes \map D) \circ  \map B \circ  \map D   =   \map B_{\rm cl} \, ,
\end{align}   
where $\map D$ is the decoherence map. It is worth stressing that a covariant broadcasting map that satisfies the classical consistency condition with respect to a single basis necessarily satisfies classical consistency with respect to \emph{every} basis (see Appendix~\ref{APXA} of the Supplemental Material for a precise statement and proof).

\begin{theo}[The Virtual Broadcasting Theorem]\label{theo:unique}
The conditions of unitary covariance, invariance under permutation of the copies, and consistency with classical broadcasting single out a unique virtual broadcasting map $\map  B$, given by
\begin{align}\label{canonical}
\map B  (\rho)     =  \frac 12     \left\{  \rho  \otimes I  ,  {\tt SWAP}\right\}   \qquad \forall \rho  \in  \Lin  (S)\, ,   
\end{align}
where $\{A,B\}:  = AB  +  BA$ denotes the anti-commutator.   
\end{theo}  

Note that Hermitian-preservation of $\map B$ is a consequence, rather than an explicit assumption, of Theorem~\ref{theo:unique} (whose proof can be found in Appendix~\ref{app:uniqueproof} of the Supplementary Material). In light of Theorem~\ref{theo:unique}, we henceforth refer to the virtual broadcasting map $\map B$ given by \eqref{canonical} as the \emph{canonical broadcasting map}. In the case of qubit systems, Ref.~\cite{HHPBS17} showed that the expression for $\map B(\rho)$ coincides with the pseudo-density matrix of Ref.~\cite{FJV15}, which captures the statistics of two-time measurements assuming trivial evolution between measurements. For arbitrary \emph{qudit} systems, the output of the canonical broadcasting map coincides with the real part of the two-point correlator of Ref.~\cite{BDOV13}. The canonical broadcasting map also appears in the construction of quantum states over time~\cite{FuPa22,FuPa22a,LiNg23}, whose relation to quantum dynamics is analogous to spacetime and its relation to classical dynamics.

In what follows, we provide  
three further results regarding the canonical broadcasting map. First, we show that the universal quantum cloning process can be characterized as the optimal physical approximation to the canonical broadcasting map. Second, we show the canonical broadcasting map may be realized as a convex combination of a \emph{virtual measure-and-prepare protocol} and the completely depolarizing channel. 
Finally, we show that the virtual broadcasting theorem yields a uniqueness result for quantum states over time. 

\medskip
{\em Optimal physical approximation to virtual broadcasting.}~ 
The universal symmetric optimal cloning process corresponds to the quantum channel $\map B^{+}:\Lin(S)\to \Lin(S_1S_2)$ given by
\begin{equation}
\map B^{+}(\rho)=\frac{2}{d+1}\Pi^+(I\otimes \rho)\Pi^+ \qquad \forall \rho\in \Lin(S),
\end{equation}
where $\Pi^{\pm}=\frac{1}{2}(I\otimes I\pm {\tt SWAP})$~\cite{BuHi96,werner-cloning,chiribella2010quantum,Harrow13}. We refer to the map $\map B^+$ as the \emph{universal cloner} for short. 

The main result of this section yields a characterization of the universal cloner as the optimal physical approximation to the canonical virtual broadcasting map. 
For quantifying this optimality we use the distance induced by the \emph{diamond norm} ~\cite{Kitaev97,Jencova14,RTG21} (which is related to the {\em completely bounded trace norm}~\cite{Paulsen02,Watrous18}), which, for any Hermitian-preserving map $\map L: \Lin(S_1)\to \Lin(S_2)$, is defined by
\begin{equation}
\label{eq:diamond-def}
    \|\map L\|_\diamond:=\max_{\omega}\|(\map I_{3}\otimes\map L)(\omega)\|_1\;,
\end{equation}
where the maximum is taken over all bipartite density matrices $\omega\in\St (S_3S_1)$, where $S_3$ has underlying Hilbert space equal to that of $S_1$, and $\|\cdot\|_1$ is the trace-norm. 

\begin{theo}
\label{theo:closest-physical}
The universal cloner is the unique quantum channel that minimizes the diamond distance to the canonical broadcasting map $\map B$; more precisely,
\begin{equation}
    \min_{\map E:\;\mathrm{channel}}\|\map B-\map E\|_\diamond=d-1\;,
\end{equation}
and the unique minimum is attained at the quantum channel $\map E=\map B^+$.
\end{theo}

The proof of Theorem~\ref{theo:closest-physical} appears in Appendix~\ref{app:closest-physical} of the Supplementary Material. It uses the spectral affine decomposition of the canonical virtual broadcasting map into two physical quantum channels given by
\begin{equation} \label{eq:cloning-decomp}
\map B=\frac{d+1}{2}\map B^{+}-\frac{d-1}{2}\map B^{-},
\end{equation}
where $\map B^{-}(\rho):=\frac{2}{d-1}\Pi^-(I\otimes \rho)\Pi^-$ for all $\rho\in \Lin(S)$. The decomposition \eqref{eq:cloning-decomp} was first proved in Ref.~\cite{BDOV13}, and the map $\map B^{-}$ may be viewed as the {\em universal anti-symmetrizer.} Note that the decomposition \eqref{eq:cloning-decomp} also yields a direct physical interpretation of the canonical virtual broadcasting map $\map B$, as it implies that $\map B$ may be simulated by a suitable post-processing of the data obtained via measurements associated with the physical processes $\map B^{\pm}$ (see
Appendix~\ref{app:spectral-decompproof} of the Supplementary Material for a proof of \eqref{eq:cloning-decomp} and additional details).

\medskip
{\em Realization of the canonical broadcasting as a virtual measure-and-prepare protocol.~}  As the canonical broadcasting map is a virtual process, in this section we extend the notion of a measure-and-prepare (M\&P) protocol to what we refer to as a  \emph{virtual} M\&P protocol. We then show that the canonical broadcasting map may be written as a convex combination of a particular virtual M\&P protocol and a physical M\&P protocol.  

Let $(M_j)_{j=1}^{k}$ be a collection of Hermitian operators satisfying the normalization condition  $ \sum_{j=1}^k    M_j    =  I $. When all the operators $M_j$ are positive semidefinite, the above collection forms a {\em positive operator-valued measure (POVM)} and represents a quantum measurement.  In the general case, we refer to the collection $(M_j)_{j=1}^{k}$  as a {\em Hermitian operator-valued measure} (HOVM), and the application of such an HOVM to a trace-one Hermitian element is referred to as a \emph{virtual measurement}. A trace-one Hermitian element $\rho  \in  \Herm_1  (S)$ is then referred to as a \emph{virtual state}. Applying an HOVM $(M_j)_{j=1}^{k}$ to a virtual state $\rho  \in  \Herm_1  (S)$ gives rise to a signed measure $\mu_\rho (j)  :=  \Tr  [M_j  \,  \rho]$ satisfying the normalization condition $\sum_j  \mu_\rho (j)   =  1$. 
 
Given an HOVM $(M_j)_{j=1}^{k}$ on an input system $S_{\rm in}$ and a collection of virtual states $(\rho_j)_{j=1}^k$ on an output system $S_{\rm out}$, one can construct a {\em virtual measure-and-prepare  (M\&P) protocol}, mathematically described by the HPTP map $\map M$ defined by sending $\rho$ to $\map M(\rho)  :  =  \sum_j  \, \Tr [M_j \rho] \,  \rho_j$.  Note that the set of all virtual M\&P protocols from  $S_{\rm in}$ to $S_{\rm out}$  is affine, and hence, in particular, convex:  for every two virtual M\&P protocols $\map M$ and $\map M'$, and for every probability $p\in  [0,1]$, their convex combination $p\,  \map M  +  (1-p) \map M'$ is also a virtual M\&P protocol.  

We now define two virtual M\&P protocols $\map M$ and $\map M'$ with a convex combination yielding the canonical broadcasting map $\map B$. The first virtual M\&P protocol $\map M$ is constructed by associating  every pure quantum state  $\psi: =  |\psi\>\<\psi| $ with a virtual state 
\begin{align}
\rho_{\psi}   :=    \frac 12   \,\left[  (d+2)  \psi    -   I \right]  
\end{align}
and by defining an HOVM with operators  $M_\psi  :=  d\, \rho_\psi $,  labelled by a continuous set of outcomes $\psi$ of rank-one projectors (see Appendix~\ref{app:HOVM} of the Supplementary Material for the precise definitions and details)
and satisfying the normalization condition $\int M_\psi \,\d \psi  =  I$, where $\d \psi$ is the normalized unitarily invariant measure. 
The virtual M\&P protocol then consists of applying the virtual measurement $M_\psi \, \d \psi$ and preparing the two-copy virtual state $\rho_\psi\otimes \rho_\psi$ conditioned on the outcome $\psi$. The overall action of this process is given by the HPTP map    
\begin{align} \label{MSCPX71}
\map M     (\rho) =  \int \Tr  [  M_\psi  \, \rho  ] \,  \rho_\psi\otimes \rho_\psi \, \d\psi  \,  \quad \forall \rho  \in \Lin(S).
\end{align}

The second virtual M\&P protocol is the \emph{completely depolarizing channel} defined by preparing two copies of the maximally mixed state, regardless of the outcome of the measurement on the input system. Namely, $\map M'$ is the channel given by 
\begin{align}\label{DSCPX87}
\map M' (\rho) = \Tr[\rho] \left(\frac{I}{d} \otimes \frac{I}{d}\right) \, \quad \forall \rho  \in \Lin(S).
\end{align}

The canonical broadcasting map $\map B$ is a convex combination of $\map M$ and $\map M'$, as illustrated by the following theorem (see Appendix~\ref{app:theoconvex} of the Supplementary Material for a proof).

\begin{theo}
\label{theo:convex}
The canonical broadcasting map $\map B$ can be decomposed as
\begin{align}\label{virtualrealization}
\map B  =    p  \,  \map M   +   (1-p) \map M'  \, , \qquad   p  :=  \frac{ 4(d+1)}{(d+2)^2} \, ,  
\end{align}
where $\map M$ is the virtual M\&P protocol given by~\eqref{MSCPX71} and $\map M'$ is the completely depolarizing channel given by~\eqref{DSCPX87}. 
\end{theo}

In light of Theorem~\ref{theo:convex}, we can think of the canonical broadcasting map  $\map B$ as realized by the following process: First, toss a biased coin with probability $p$ of getting heads. If the outcome is heads, perform the HOVM $\{M_\psi\}_{\psi}$ and prepare the virtual state $\rho_\psi\otimes\rho_\psi$ if the outcome of the measurement is $\psi$ (since the probability that the outcome is exactly  $\psi$ is zero, and similarly for the preparation of two copies of $\rho_\psi$, the rigorous meaning here is captured by the integral expression in Eq.~\eqref{MSCPX71}). If the outcome is tails, prepare two copies of the maximally mixed state.   Note that, since the set of virtual M\&P protocols is convex, we can think of the above randomization as a single M\&P protocol.  

\medskip
{\em Canonical states over time from virtual broadcasting.~}  
If $\map B$ is a broadcasting map and $\mathcal{E}:\Lin(S_1)\to \Lin(S_2)$ is a quantum channel, then for every state $\rho\in \St(S_1)$ the element $\mathcal{E}\star \rho\in \Lin(S_1S_2)$ given by
\begin{equation} \label{CSTXS971}
\mathcal{E}\star \rho=(\mathcal{I}\otimes \mathcal{E})\left(\mathcal{B}(\rho)\right)
\end{equation}
is an example of a quantum state over time associated with the process of $\rho$ evolving under the channel $\mathcal{E}$. More generally~\cite{HHPBS17,FuPa22,FuPa22a}, a \emph{quantum state over time} is an element $\mathcal{E}\star \rho\in \Lin(S_1S_2)$ such that
\begin{equation} \label{MCXD67}
\Tr_{S_2}[\map E\star\rho]=\rho \quad\text{and}\quad \Tr_{S_1}[\map E\star\rho]=\map E(\rho).
\end{equation}
As $\map E$ and $\rho$ vary, the assignment $\star:(\mathcal{E},\rho)\mapsto \mathcal{E}\star \rho\in \Lin(S_1S_2)$ is then referred to as a \emph{state over time function}. A state over time function given by~\eqref{CSTXS971} for some broadcasting map $\map B$ is said to satisfy the \emph{broadcasting condition}. This condition can be derived as a consequence of consistency with probabilistic mixtures of states and consistency with post-processing via arbitrary quantum channels and measurements (see Appendix~\ref{app:broadcastingcondition} of the Supplementary Material for details and an operational derivation of the broadcasting condition).

A quantum state over time encodes \emph{spatio-temporal} correlations that result in the process of a state evolving according to a quantum channel, and it serves as a quantum analog of a joint probability distribution~\cite{HHPBS17,FuPa22}. Unlike quantum states at a fixed point in time, quantum states over time are not positive in general, indicating a further analogy with the Lorentzian signature of spacetime, as opposed to the Riemannian signature of a spacelike hypersurface~\cite{ADM59,ADM08}. 

An axiomatic study of state over time functions was initiated in Ref.~\cite{HHPBS17}, and Ref.~\cite{FuPa22} used the canonical broadcasting map in formula~\eqref{CSTXS971} to give the first example of a state over time function satisfying the axioms put forth in Ref.~\cite{HHPBS17}. While Ref.~\cite{LiNg23} has recently proved that the state over time function constructed in Ref.~\cite{FuPa22} is uniquely characterized by a different list of axioms, we provide an alternative characterization as a direct corollary of our virtual broadcasting theorem. For this, we formulate three conditions for state over time functions, which under the assumption of the broadcasting condition \eqref{CSTXS971}, are in fact equivalent to the conditions appearing in the virtual broadcasting theorem.  

First, a state over time function $\star$ is said to be \emph{covariant} whenever 
\begin{equation}
(\mathcal{U}\otimes \mathcal{V})(\mathcal{E}\star \rho)=\mathcal{E}'\star \mathcal{U}(\rho) \qquad \forall \rho\in \St(S_1)
\end{equation}
for all pairs of quantum channels $\mathcal{E},\mathcal{E}'$ satisfying $\mathcal{V}\circ\mathcal{E}=\mathcal{E}'\circ\mathcal{U}$ for some unitary channels $\mathcal{U}$ and $\mathcal{V}$. Second, a state over time function $\star$ is said to be \emph{permutation invariant} whenever
\begin{equation}
{\tt SWAP}(\mathcal{I}\star \rho){\tt{SWAP}}=\mathcal{I}\star \rho \qquad \forall \rho\in \St(S_1).
\end{equation}
Finally, a state over time function $\star$ is said to be \emph{classically consistent} whenever 
\begin{equation}
(\map D\otimes\map D')\big(\map E\star\map D(\rho)\big)=
\map E\star_{\rm cl} \rho,
\end{equation}
where $\map D$ and $\map D'$ are the decoherence maps on the domain and codomain of $\map E$, respectively,  $\map E\star_{\rm cl} \rho := (\map I\otimes\map E)\big(\map B_{\rm cl}(\rho)\big)$, and $\map E$ is a  channel satisfying $\map D'\circ\map E=\map E\circ\map D$ (so that $\map E$ commutes with the application of decoherence maps on the input and outputs). 

Here, the permutation invariance condition coincides with the ``time-reversal symmetry'' axiom of Ref.~\cite{LiNg23}, while the covariance and classical consistency conditions do not appear as axioms in Ref.~\cite{LiNg23}.  Requiring these conditions singles out a unique state over time function:

\begin{theo}
\label{thm:SOTcharacterization}
Let $\star$ be a state over time function satisfying the broadcasting condition~(\ref{CSTXS971}).  If $\star$ is covariant, permutation invariant, and classically consistent, then  
\begin{equation}
\label{eq:SOTbcastthm}
\mathcal{E}\star \rho=(\mathcal{I}\otimes \mathcal{E})(\map B(\rho)) \quad \quad \forall (\mathcal{E},\rho),
\end{equation}
where $\map B$ is the canonical broadcasting map.
\end{theo} 

\medskip

\noindent
{\bf Proof of Theorem~\ref{thm:SOTcharacterization} }
Since $\star$ satisfies the broadcasting condition \eqref{CSTXS971}, the associated broadcasting map $\map B$ is given by $\map B(\rho)=\map I\star \rho$. Moreover, $\star$ is covariant, permutation invariant, and classically consistent, so that $\map B$ is as well. Hence, Theorem~\ref{theo:unique} implies~\eqref{eq:SOTbcastthm}. 
\qed

\medskip
{\em Conclusions.~} 
In this Letter, we have proved the \emph{virtual broadcasting theorem}, which states that a unique Hermitian-preserving broadcasting map is singled out by the natural conditions of covariance, permutation invariance, and consistency with classical broadcasting. While not a genuine physical process, the canonical virtual broadcasting map is an affine combination of the physical processes corresponding to the universal quantum cloner and the universal quantum anti-symmetrizer. Moreover, we showed that the universal quantum cloner is the optimal physical approximation to our canonical broadcasting map with respect to the diamond norm. We also showed that the canonical virtual broadcasting map is a convex combination of a virtual measure-and-prepare protocol and the completely depolarizing channel. Finally, we extended the conditions of the virtual broadcasting theorem to analogous conditions for quantum states over time, thus yielding a uniqueness result for state over time functions under minimal assumptions.

\vspace{3mm}
{\bf Acknowledgements}
A.J.P.\ and J.F.\ acknowledge support from the Blaumann Foundation and thank Seok Hyung Lie for discussions. 
A.J.P.\ and F.B.\ acknowledge support from MEXT-JSPS Grant-in-Aid for Transformative Research Areas (A) ``Extreme Universe'', No.\ 21H05183.   
F.B.\ acknowledges
support from MEXT Quantum Leap Flagship Program (MEXT QLEAP)
Grant No. JPMXS0120319794; from JSPS KAKENHI Grants  No. 20K03746  and No. 23K03230. 
G.C.\ acknowledges support from the Hong Kong Research Grant Council  through the Senior Research Fellowship Scheme SRFS2021-7S02 and through grant no. 17307520, and from the John Templeton Foundation through grant 62312, The Quantum Information Structure of Spacetime (qiss.fr).   
The opinions expressed in this publication are those of the authors and do not necessarily reflect the views of the John Templeton Foundation. Research at the Perimeter Institute is supported by the Government of Canada through the Department of Innovation, Science and Economic Development Canada and by the Province of Ontario through the Ministry of Research, Innovation and Science. 
\appendix

\section{Basis independence of classical consistency} \label{APXA}

In this section we show that for any covariant broadcasting map $\map B$, the notion of consistency with classical broadcasting is independent of the basis used to formulate the notion. The precise statement is the following: 
\begin{prop}
\label{prop:anybasis}
Let $\{|i\>\}_{i=1}^{d}$ and $\{U|i\>\}_{i=1}^{d}$ be two orthonormal bases for a quantum system $S$, where $U$ is a unitary operator, and let $\map D$ and $\map D'$ be their corresponding decoherence maps. Then a covariant broadcasting map $\map B$ satisfies the classical consistency condition 
\begin{equation}
\label{eq:classconsistcondoriginal}
(\map D\otimes\map D)\circ\map B\circ\map D=\map B_{\rm cl}
\end{equation}
if and only if it satisfies 
\begin{equation}
\label{eq:classconsistcondprime}
(\map D' \otimes\map D')\circ\map B\circ\map D'=\map B'_{\rm cl},
\end{equation}
where $\map B'_{\rm cl}$ is the classical broadcasting map uniquely determined by 
\begin{equation}
\map B'_{\rm cl}(U|i\>\<j|U^{\dag})=\delta_{ij}U|i\>\<i|U^{\dag}\otimes U|i\>\<i|U^{\dag}
\end{equation}
for all $i,j\in\{1,\dots,d\}$. 
\end{prop}

\noindent
{\bf Proof of Proposition~\ref{prop:anybasis}}
First, note that 
\begin{align}
\map B_{\rm cl}
&=(\map D\otimes\map I)\circ\mu^{*}\circ\map D
=(\map I\otimes\map D)\circ\mu^{*}\circ\map D\nonumber\\
&=(\map D\otimes\map D)\circ\mu^{*}\circ\map D,
\label{eq:classicalbcastidentity}
\end{align}
where $\mu^*$ is the Hilbert--Schmidt adjoint of the multiplication map $\mu:\Lin(S_1)\otimes\Lin(S_2)\to\Lin(S)$, the latter of which is uniquely determined by $\mu(A_1 \otimes A_2)=A_1 A_2$ for all $A_1\in\Lin(S_1)$, $A_2\in\Lin(S_2)$. Indeed, temporarily writing $\Psi:=(\map D\otimes\map I)\circ\mu^{*}\circ\map D$, one finds
\begingroup
\allowdisplaybreaks
\begin{align}
\Psi(|i\>\<j|)&=(\map D\otimes \map I)\big(\mu^*(\delta_{ij}|i\>\<i|)\big)\nonumber\\
&=\delta_{ij}(\map D\otimes \map I)\left(\sum_{k=1}^{d}|i\>\<k|\otimes|k\>\<i|\right)\nonumber\\
&=\delta_{ij}\sum_{k=1}^{d}\delta_{ik}|i\>\<i|\otimes|i\>\<i|\nonumber\\
&=\delta_{ij}|i\>\<i|\otimes|i\>\<i|
=\map B_{\rm cl}(|i\>\<j|),
\end{align}
\endgroup
where the second equality follows from \cite[Lemma~3.39]{PaRuBayes}. A similar calculation shows the other identities in~\eqref{eq:classicalbcastidentity}. 

Second, note the relations  
\begin{equation}
\label{eq:covardecoh}
\map D'=\map U\circ\map D\circ\map U^{-1}
\end{equation}
and  
\begin{equation}
\label{eq:covarmult}
\map U^{-1}\circ\mu\circ(\map U\otimes\map U)=\mu,
\end{equation}
where $\map U$ is defined by $\map U(A)=UAU^{\dag}$ for all inputs $A$.
The former is an immediate consequence of the definitions and the latter follows from setting $\Phi:=\map U^{-1}\circ\mu\circ(\map U\otimes\map U)$ and verifying by the computation  
\begin{align}
\Phi(A_1\otimes A_2)&=\map U^{-1}\big(\mu(UA_{1}U^{\dag}\otimes UA_{2}U^{\dag})\big)\nonumber\\
&=\map U^{-1}(UA_{1}A_{2}U^{\dag})\nonumber\\
&=A_{1}A_{2}=\mu(A_{1}\otimes A_{2}).
\label{eq:proofofcovmu}
\end{align}
Therefore, if $\map B$ satisfies the classical consistency condition~\eqref{eq:classconsistcondoriginal}, then 
\begingroup
\allowdisplaybreaks
\begin{align}
&(\map D'\otimes\map D')\circ\map B\circ\map D' \nonumber\\
&=(\map U\otimes\map U)\circ(\map D\otimes\map D)\circ(\map U^{-1}\otimes\map U^{-1})\circ\map B\circ \map U\circ\map D\circ\map U^{-1}\nonumber\\
&=(\map U\otimes\map U)\circ(\map D\otimes \map D)\circ\map B\circ\map D\circ\map U^{-1}\nonumber\\
&=(\map U\otimes\map U)\circ\map B_{\rm cl}\circ \map U^{-1}\nonumber\\
&=\map B'_{\rm cl}, 
\end{align}
\endgroup
where the first equality follows from the interchange law of $\otimes$ and $\circ$, the second equality follows from covariance of the broadcasting map $\map B$, the third equality follows from the assumption~\eqref{eq:classconsistcondoriginal}, and the last equality follows from~\eqref{eq:covardecoh},~\eqref{eq:covarmult}, and the definition of $\map B'_{\rm cl}$.  
Thus,~\eqref{eq:classconsistcondprime} holds. 
A similar calculation shows the converse, namely if $\map B$ satisfies the condition~\eqref{eq:classconsistcondprime}, then it satisfies the classical consistency condition~\eqref{eq:classconsistcondoriginal}. 
\qed

\section{A one-parameter family of virtual broadcasting maps} \label{APX0}

Given $\lambda\in \R$, let $\map B_{\lambda}:\Lin(S)\to \Lin(S_1S_2)$ be the map given by
\begin{equation}
\map B_{\lambda}(\rho):= \frac{1}{2}\left\{\rho  \otimes I,{\tt SWAP}\right\}+i\lambda \left[\rho\otimes I, {\tt SWAP}\right]  ,
\end{equation}
where $i:=\sqrt{-1}$ and $[A,B]:=AB-BA$ denotes the commutator. 

\begin{prop}
\label{prop:family}
The map $\map B_{\lambda}$ is a virtual broadcasting map that is covariant and classically consistent for all $\lambda \in \R$. Furthermore, $\map B_{\lambda}$ is invariant under permutation of copies if and only if $\lambda=0$.  
\end{prop}

We first prove two lemmas.

\begin{lemma}
\label{lemma:MX}
Let $\map U:\Lin(S)\to \Lin(S)$ be the map given by $\map U(A)=UAU^{\dag}$ for a unitary $U\in \Lin(S)$, and let $\mu:\Lin(S_1S_2)\to \Lin(S)$ and $\tilde{\mu}:\Lin(S_1S_2)\to \Lin(S)$ be the maps corresponding to the unique linear extensions of the assignments $\mu(A_{1}\otimes A_{2})=A_{1}A_{2}$ and $\tilde{\mu}(A_{1}\otimes A_{2})=A_{2}A_{1}$. Then
\begin{equation} \label{MX1}
\map U\circ \mu \circ (\map U^{-1}\otimes \map U^{-1})=\mu
\end{equation}
and
\begin{equation} \label{MX2}
\map U\circ \tilde{\mu} \circ (\map U^{-1}\otimes \map U^{-1})=\tilde{\mu}.
\end{equation}
\end{lemma}

\noindent
{\bf Proof of Lemma~\ref{lemma:MX}}
Eq.~\eqref{MX1} is equivalent to Eq.~\eqref{eq:covarmult}, which was proved in~\ref{eq:proofofcovmu}. The proof of~\eqref{MX2} is similar.
\qed

\begin{lemma}
\label{lemma:DCX}
Let $\map D$ be a decoherence map with respect to some orthonormal basis. Then $\map D=\map D^*$, where $\map D^*$ denotes the Hilbert--Schmidt adjoint of $\map D$. 
\end{lemma}

\noindent
{\bf Proof of Lemma~\ref{lemma:DCX}} 
The rank-one projectors $\{|i\>\<i|\}$ form a set of Kraus operators for $\map D$, i.e.,  $\map D(\rho)=\sum_{i=1}^{d}|i\>\<i| \rho |i\>\<i|$ for all states $\rho$. Since the Kraus operators are self-adjoint, and because the Hilbert--Schmidt adjoint of a map is given by the adjoint of the Kraus operators, $\map D^{*}=\map D$. \qed

\noindent
{\bf Proof of Proposition~\ref{prop:family}}
We first show that $\map B_{\lambda}$ is a virtual broadcasting map for all $\lambda\in \R$. For this, let $\{|i\>\}_{i=1}^{d}$ be an orthomoral basis of $S$. Then
\begin{align}
\Tr_{S_1}\big[(\rho\otimes I){\tt SWAP}\big]&=\Tr_{S_1}\left[\sum_{i,j}\rho |i\>\<j|\otimes |j\>\<i|\right] \nonumber \\
&=\sum_{i,j}\<j|\rho|i\>|j\>\<i|\nonumber\\
&=\sum_{i,j}|j\>\<j|\rho|i\>\<i|=\rho
\end{align}
for all $\rho\in \Lin(S)$. 
A similar calculation yields $\Tr_{S_2}\big[(\rho\otimes I){\tt SWAP}\big]=\rho$, from which it follows that $\map B_{\lambda}$ satisfies the broadcasting condition for all $\lambda\in \R$. To show $\map B_{\lambda}$ is Hermitian-preserving, we prove that $\map B_{\lambda}$ is $\dag$-preserving, i.e., $\map B_{\lambda}(A^{\dag})=\map B_{\lambda}(A)^{\dag}$ for all $A\in\Lin(S)$, which is equivalent. For this, we first note that ${\tt SWAP}^{\dag}={\tt SWAP}$, which immediate follows from the formula ${\tt SWAP}=\sum_{i,j} |i\>\<j|\otimes |j\>\<i|$. For all $A\in \Lin(S)$, we then have
\begin{align}
\frac{1}{2}\left\{(A^{\dag}\otimes I),{\tt SWAP}\right\}&=\frac{1}{2}\left\{(A^{\dag}\otimes I),{\tt SWAP}^{\dag}\right\} \nonumber\\
&=\frac{1}{2}\left\{(A\otimes I),{\tt SWAP}\right\}^{\dag},
\end{align}
and
\begin{align}
\lambda i\left[(A^{\dag}\otimes I),{\tt SWAP}\right]&=\lambda i\left[(A^{\dag}\otimes I),{\tt SWAP}^{\dag}\right] \nonumber \\
&=\left(-\lambda i\left[{\tt SWAP},(A\otimes I)\right]\right)^{\dag} \nonumber \\
&=\left(\lambda i\left[(A\otimes I){\tt SWAP}\right]\right)^{\dag},
\end{align}
from which it follows that $\map B_{\lambda}(A^{\dag})=\map B_{\lambda}(A)^{\dag}$, as desired.

Now let $U\in \Lin(S)$ be unitary. To prove $\map B_{\lambda}$ is covariant, note that
$\map B_{\lambda}$ is covariant if and only if 
\begin{equation}\label{EQX17}
\map U \circ \map B_{\lambda}^*\circ (\map U^{-1}\otimes \map U^{-1})=\map B_{\lambda}^*,
\end{equation}
where $\map B_{\lambda}^*$ is the Hilbert--Schmidt adjoint of $\map B_{\lambda}$. Moreover, since
\begin{equation}\label{HSABX57}
\map B_{\lambda}^*=\frac{1}{2}(\mu+\tilde{\mu})+i\overline{\lambda}(\mu-\tilde{\mu}),
\end{equation}
where $\mu$ and $\tilde{\mu}$ are as in Lemma~\ref{lemma:MX},
Eqs. \eqref{MX1} and \eqref{MX2} imply that \eqref{EQX17} indeed holds. Hence, $\map B_{\lambda}$ is covariant.

To prove $\map B_{\lambda}$ is classically consistent, let $\map D$ be a decoherence map with respect to some orthonormal basis. By \eqref{eq:classicalbcastidentity}, $\map B_{\lambda}$ is classically consistent if and only if
\begin{equation} \label{CCX87}
\map D\circ \map B_{\lambda}^*\circ (\map D\otimes \map D)=\map D\circ \map \mu\circ (\map D\otimes \map D)
\end{equation}
since $\map D^*=\map D$ by Lemma~\ref{lemma:DCX}. 
Moreover, it follows from \eqref{HSABX57} that
\begin{equation}
[A,B]=0\implies \map B_{\lambda}^*(A\otimes B)=\mu(A\otimes B),
\end{equation}
and since $[\map D(A), \map D(B)]=0$ for all $A$ and $B$, it follows that \eqref{CCX87} indeed holds.

Now we show $\map B_{\lambda}$ is invariant under permutation of copies if and only if $\lambda=0$. First, note that $\map B_{\lambda}$ is invariant under permutation of copies if and only if $\map B_{\lambda}=\gamma \circ \map B_{\lambda}$, where $\gamma$ is the lexicographic swap isomorphism given by
\begin{equation}
\gamma(\rho)={\tt SWAP}\,\rho \,{\tt SWAP}
\end{equation}
for all $\rho\in \Lin(S)$. Now if $\lambda=0$, then it follows from \eqref{HSABX57} that $\map B_{0}$ may be written as
\begin{equation}
\map B_{0}=\frac{1}{2}(\mu^*+\gamma\circ \mu^*),
\end{equation}
which, together with the fact that $\gamma\circ \gamma =\map I$, implies $\map B_{0}=\gamma \circ \map B_{0}$. Thus, $\map B_{0}$ is invariant under permutation of copies. 

For the converse, suppose $\map B_{\lambda}$ is invariant under permutation of copies. Since $\map B_{\lambda}$ may be written as
\begin{equation}
\map B_{\lambda}=\frac{1}{2}(\mu^*+\gamma\circ \mu^*)+i\lambda(\mu^*-\gamma\circ \mu^*),
\end{equation}
the equation $\map B_{\lambda}=\gamma \circ \map B_{\lambda}$ implies
\begin{equation}
i\lambda(\mu^*-\gamma\circ \mu^*)=i\lambda(\gamma\circ \mu^*-\mu^*),
\end{equation}
which is equivalent to the condition
\begin{equation}
\lambda[A,B]=\lambda[B,A]
\end{equation}
for all $A,B$, which holds if and only if $\lambda=0$ for a quantum system $S$ of dimension $d>1$. \qed

\section{Proof of Theorem \ref{theo:unique}}
\label{app:uniqueproof}

Let $\map B$ be a broadcasting map that satisfies covariance under unitary evolution, invariance under permutations of the copies, and consistency with classical broadcasting. The classical consistency condition implies that the broadcasting map $\map B$ acting on a diagonal pure state, after the action of decoherence on its outputs, should coincide with the unique classical broadcasting map $\map B_{\rm cl}$, i.e.,  \begin{align} 
&|i\>\<i|  \otimes |i\>\<i| =(\map D  \otimes \map D )\,  \map B (|i\>\<i| )\nonumber\\
&=\sum_{j,k}\<j|\<k|\map B(|i\>\<i|)|j\>|k\>\;|j\>\<j|\otimes|k\>\<k|\;\;\, \forall i\in    [d] \, ,
\end{align}
where $[d]:=\{1,\dots,d\}$. 
By linear independence of the basis $\{|j\>\<k|\otimes|m\>\<n|\}$,  this implies
\begin{equation}
\<j| \<k|   \,   \map B    (|i\>\<i|  )  \,|j\>|k\>  = \delta_{ij}\,  \delta_{ik}\,,  \qquad\forall i,j,k \in  [d] \, .
\end{equation}
Since $\map B$ is trace-preserving by the broadcasting condition, $\map B(|i\>\<i|)$ is a trace-1 operator. Hence, the above equality implies that $\map B(|i\>\<i|)$ can be written as 
\begin{align}\label{Oi}
\map B (|i\>\<i|)   =   |i\>\<i|  \otimes |i\>\<i|    +   O_i  \, ,
\end{align}  
where $O_i$ is an operator satisfying 
  \begin{align}\label{Oiproperty}
\<j|\<k|  \, O_i \,  |j\>|k\>  =  0  \qquad \forall  j,k \in  [d] \, .
  \end{align}  

The following lemma determines the form of the operator $O_i$. 
\begin{lemma}\label{lem:Oiswap}
The operator $O_i$ defined in Eq. (\ref{Oi}), for a covariant and classically consistent broadcasting map $\map B$, is of the form  
\begin{equation}\label{Oiswap}
O_i   =  \lambda_i \,     {\tt SWAP}   \,  (    |i\>\<i|  \otimes P_\perp)     +   \nu_i  \,  (    |i\>\<i|  \otimes P_\perp) \, {\tt SWAP} \, ,
\end{equation}
where $\lambda_i,\nu_i$ are complex numbers and $P_\perp  : =   I-  |i\>\<i|   =    \sum_{j\not  = i} |j\>\<j|$. 
\end{lemma}
\Proof  
Let $\grp U_i$ be the set of all unitary operators $U$ of the form  $U =  |i\>\<i|  \oplus   V $, with $V$ an arbitrary  unitary operator acting on the subspace $\spc H_\perp : = P_{\perp}\spc H\equiv  \Span\{|j\> ~|~   j\not  = i\}$, where $\spc H$ is the underlying Hilbert space of the quantum system.   Since the operator $U$ satisfies the condition  $U|i\>  =  |i\>$, we have
       \begin{align}
\nonumber |i\>\<i|&\otimes|i\>\<i|    +  O_i    =         \map B  (  |i\>\<i|  ) \\
  \nonumber  &  = \map B  (U  |i\>\<i|  U^\dag) \\
  \nonumber  &     =    (U\otimes U)   \map B  (  |i\>\<i|)  (U \otimes U)^\dag \\
    \nonumber   &  =   (U\otimes U)   \,  (  |i\>\<i|  \otimes |i\>\<i|    +   O_i ) \,   (U \otimes U)^\dag    \\
      &  =    |i\>\<i|  \otimes |i\>\<i|    +  (U\otimes U)   \,    O_i  \,   (U \otimes U)^\dag   \, ,  \end{align}
      where the third equality follows from the covariance of the map $\map B$.

 The above equality is equivalent to the fact that the operator $O_i$ satisfies the commutation relation    
      \begin{align}\label{Oicommutes}
[O_i,  U\otimes U]    =  0  \qquad\forall   U  \in  \grp U_i\, .
      \end{align}
The implications of this commutation relation can be worked out explicitly using the representation theory of the unitary group.   The operators  
\begin{align}
\nonumber  U\otimes U   &=   ( |i\>\<i|  \otimes |i\>\<i| ) \; \oplus \;  (   |i\>\<i|  \otimes V ) \nonumber \\
&\quad\;\oplus  ( V\otimes  |i\>\<i|) \;   \oplus \;  (V\otimes V)  
\end{align} 
form a reducible unitary representation of the group $\grp  U(d-1)$ on $\spc H\otimes \spc H$.      This representation has five orthogonal irreducible subspaces inside $\spc H\otimes\spc H$, namely 
\begin{align}
\nonumber \spc H_1  &:  =  \Span  \{  |i\>|i\>\}\\
\nonumber \spc H_{2}  &:=\Span\{  |i\>|\phi\>  ~|~ |\phi\>  \in \spc H_\perp \} \\
\nonumber \spc H_{3}  &:=\Span\{  |\phi\>|i\>  ~|~ |\phi\>  \in \spc H_\perp \} \\
\nonumber \spc H_{4}  &:=  \Span\{   |\phi\>|\phi\>   ~|~  |\phi\>  \in \spc H_\perp \}  \\ 
\spc H_{5}  &:=   \Span\{   \Pi^-   |\phi\>|\psi\>  ~|~ |\phi\>  \in  \spc H_\perp \, , |\psi\>  \in  \spc H_\perp  \} \, ,
\end{align}
where $\Pi^-  :  =  \frac{1}{2}(I\otimes I  - {\tt SWAP}) $ is the projector onto the antisymmetric subspace  of $\spc H\otimes\spc H$, which also restricts to the projector onto the antisymmetric subspace of $\spc H_{\perp}\otimes\spc H_{\perp}$ (the four invariant subspaces $\spc H_1$, $\spc H_2$, $\spc H_3$, and $\spc H_4\oplus \spc H_5$ inside $\spc H\otimes \spc H$ correspond, respectively, to the actions obtained from the $(|i\>\<i|\otimes|i\>\<i|)$, $(|i\>\<i| \otimes V )$, $( V\otimes  |i\>\<i|)$, and $(V\otimes V)$ direct sum components of $U\otimes U$). 
By Schur's lemma~\cite{Hall13,Wo17}, the commutation relation (\ref{Oicommutes}) implies that  $O_i$ must be a linear combination of the unitary equivalences $T_{lk}  :  \spc H_k  \to \spc H_l$, provided they exist. 
Furthermore, for $k=l$, the isomorphism $T_{kk}$ is  just the identity, or equivalently the projector onto the subspace $\spc H_k$ when viewed as an operator on $\spc H\otimes \spc H$. Explicitly, these projectors are given by 
\begin{align}
\nonumber  T_{11}  &= |i\>\<i| \otimes |i\>\<i|\\ 
\nonumber T_{22}   &=  |i\>\<i|  \otimes P_\perp  \\
\nonumber T_{33}&=    P_\perp \otimes |i\>\<i|  \\
T_{44}  +  T_{55}  &=  P_\perp \otimes P_\perp \, .     \label{relazioni} 
\end{align}  For $k\not  = l$, the only values of $\{k,l\}$ corresponding to equivalent representations are $\{2,3\}$.  The two intertwiners between the subspaces $\spc H_2$ and $\spc H_3$ are 
\begin{align}\label{isomorphisms}
\nonumber T_{32}  & =   \sum_{j\not =  i}   \,  |j\>\<i|  \otimes |i\>\<j|     =  {\tt SWAP}   \,  (  |i\>\<i|  \otimes P_\perp)     \\
  T_{23}  &=  T_{32}^\dag  =     (  |i\>\<i|  \otimes P_\perp)      \,  {\tt SWAP}    \, 
\end{align} 
since $T_{32}(|i\>|\phi\>)=|\phi\>|i\>$ and $T_{23}(|\phi\>|i\>)=|i\>|\phi\>$ for all $|\phi\>\in\spc H_{\perp}$. 
Hence, we must have 
      \begin{align}
      O_i   = & \sum_{k=1}  a_{kk}\,  T_{kk}    +   a_{23}  \,  T_{23}   +   a_{32}\,  T_{32} \, ,
      \end{align}
    for some set of complex coefficients $\{a_{kl}\}$. Combining the relations (\ref{relazioni}) with Eq. (\ref{Oiproperty}), we obtain   
   \begin{align}
\nonumber a_{11}   &=  \<i|  \<i|   O_i  \, |i\>|i\>  =  0\\
\nonumber a_{22}  &  =  \<i|  \<j|  O_i\,  |i\>|j\>   =   0  \qquad \forall j\not =  i \\
\nonumber a_{33}  &  =  \<j|  \<i|  O_i\,  |j\>|i\>   =   0  \qquad \forall j\not =  i \\
\nonumber   a_{44}  &  =  \<j|  \<j|  O_i\,  |j\>|j\>   =   0  \qquad \forall j\not =  i \\
a_{55}    &  =  \<j|  \<k|  O_i\,  |j\>|k\>   =   0  \qquad \forall j,k\not =  i  \, .
   \end{align}    
    
   Defining $\lambda_i:  = a_{32}$ and $\nu_i:=a_{23}$, and using Eq. (\ref{isomorphisms}), we then obtain Eq. (\ref{Oiswap}). \qed 
 
   \medskip  
   We now proceed to proving Theorem~\ref{theo:unique}. Using Lemma~\ref{lem:Oiswap}, Eq.~\eqref{Oi}, and assuming invariance under permutation of the copies for $\map B$, we obtain 
   \begin{align}
   \lambda_{i}T_{32}+\nu_{i}T_{23}&=O_{i}\nonumber \\
   &={\tt SWAP}\,O_{i}\,{\tt SWAP}\nonumber \\
   &=\lambda_{i}T_{23}+\nu_{i}T_{32},
   \end{align}
   where we used the notation from Eq.~\ref{isomorphisms}. Hence, $\nu_{i}=\lambda_{i}$.
   
 Now, combining Lemma~\ref{lem:Oiswap} with Eq.~(\ref{Oi}), we then obtain  
 \begin{align}
\nonumber  \map B (|i\>\<i|)      &=  |i\>\<i|  \otimes |i\>\<i| +   \lambda_i \,     {\tt SWAP}   \,  (    |i\>\<i|  \otimes P_\perp) \\
\nonumber  & \quad    +  \lambda_i  \,  (    |i\>\<i|  \otimes P_\perp) \, {\tt SWAP}  \\
 \nonumber     &  =    (1-2\lambda_i)~   |i\>\<i|  \otimes |i\>\<i| +   \lambda_i \,     {\tt SWAP}   \,  (    |i\>\<i|  \otimes  I)   \\
  & \quad  +   \lambda_i  \,  (    |i\>\<i|  \otimes I) \, {\tt SWAP}  \, .
  \end{align}  
 Now, let $|\psi\>\in \spc H$ be an arbitrary unit vector, and let  $W$ be a unitary operator such that $W|i\>  =  |\psi\>$.  The covariance of $\map B$ implies 
 \begin{align}
\nonumber  \map B &(|\psi\>\<\psi|)      =   \map B (  W  |i\>\<i|  W^\dag)  \\
 \nonumber  &    =  (W\otimes W)  \,  \map B (|i\>\<i|) \,  (W\otimes W)^\dag  \\
 \nonumber    &  =    (1-2\lambda_i )~   |\psi\>\<\psi|  \otimes |\psi\>\<\psi| +   \lambda_i \,     {\tt SWAP}   \,  (    |\psi\>\<\psi|  \otimes  I) \\
  & \quad    +  \lambda_i  \,  (    |\psi\>\<\psi|  \otimes I) \, {\tt SWAP}  \, 
   \end{align}   
   because $[W\otimes W,{\tt SWAP}]=0=[W^{\dag}\otimes W^{\dag},{\tt SWAP}]$. 
 Equivalently, the above equation reads
 \begin{align}
 \nonumber & \map B (|\psi\>\<\psi|)    -   \lambda_i \,     {\tt SWAP}   \,  (    |\psi\>\<\psi|  \otimes  I)     -  \lambda_i  \,  (    |\psi\>\<\psi|  \otimes I) \, {\tt SWAP} \\
 &  =  (1-2\lambda_i)~   |\psi\>\<\psi|  \otimes |\psi\>\<\psi|   \, .   
 \end{align}
 Recall that $|\psi\>$ is an arbitrary unit vector.  For the above equation to hold for every $|\psi\>$, one must have  $2\lambda_i =  1$ in order for $\map B$ to be a linear transformation.  
 Therefore, $\lambda_{i}=1/2$, and we obtain
 \begin{align}
\nonumber \map B (|\psi\>\<\psi|)  & =     \frac 12    {\tt SWAP}   \,  (    |\psi\>\<\psi|  \otimes  I)     +    \frac 12  \,  (    |\psi\>\<\psi|  \otimes I) \, {\tt SWAP}  \\
&=  \frac 12  \bigg\{  {\tt SWAP}   \, ,      |\psi\>\<\psi|  \otimes  I       \bigg\} \quad \forall  |\psi\>  \in  \spc H\, .
 \end{align}   
Since the map $\map B$ is linear and  the projectors  $|\psi\>\<\psi|$  form a spanning set for $\Lin  (S)$, this identity concludes the proof of Theorem \ref{theo:unique}. \qed

\section{Proof of Equation~\eqref{eq:cloning-decomp}}
\label{app:spectral-decompproof}

The Choi operator~\cite{Ch75} of the canonical broadcasting map~\eqref{canonical} $\map B$ is given by
\begin{equation}
    C(\map B):=({\map B}\otimes \map I)(\Omega)\,, 
\end{equation}
where $\Omega:=\sum_{i,j}|ii\>\<jj|\equiv\sum_{i,j}|i\>\<j|\otimes|i\>\<j|$. This can be explicitly computed as
\begin{align}
C(\map B)&=\sum_{i,j}\frac{1}{2}\big\{|i\>\<j|\otimes I,{\tt SWAP}\big\}\otimes|i\>\<j|\nonumber\\
&=\frac{1}{2}\sum_{i,j}\Big(\big((|i\>\<j|\otimes I){\tt SWAP}\big)\otimes|i\>\<j|\nonumber\\
&\qquad\quad+\big({\tt SWAP}(|i\>\<j|\otimes I)\big)\otimes|i\>\<j|\Big) \nonumber\\
&=\frac{1}{2}\sum_{i,j}\Big(\big({\tt SWAP}(I\otimes |i\>\<j|)\big)\otimes|i\>\<j|\nonumber\\
&\qquad\quad+\big((I \otimes |i\>\<j|){\tt SWAP}\big)\otimes|i\>\<j|\Big)\nonumber\\
&=\frac{1}{2}\sum_{i,j}\Big(({\tt SWAP}_{12}\otimes I_{3})(I_{1}\otimes|ii\>\<jj|)\nonumber\\
&\qquad\quad+(I_{1}\otimes|ii\>\<jj|)({\tt SWAP}_{12}\otimes I_{3})\Big)\nonumber\\
    &=\frac 12\big\{\texttt{SWAP}_{12}\otimes I_{3},I_{1}\otimes\Omega_{23}\big\}\,,
    \label{eq:ChoiCB}
\end{align}
where the property $(A\otimes B){\tt SWAP}={\tt SWAP} (B\otimes A)$ was used in the third equality. 
In the above equation, indices have been included to identify the appropriate tensor factor for additional clarity; namely, spaces labeled ``1'' and ``2'' represent the output of $\map B$, while ``3'' represents the input.
It then follows that
\begin{equation}\label{CHDCX77}
C(\map B)=\hat B^+-\hat B^-,
\end{equation}
where 
\begin{equation} \label{HXB87}
\hat B^{\pm}=(\Pi^{\pm}\otimes I_{3})(I_{1}\otimes\Omega_{23})(\Pi^{\pm}\otimes I_{3})
\end{equation}
and $\Pi^{\pm}=\frac 12(I\pm\texttt{SWAP})$. 
This already shows that $C(\map B)$ is a difference of two positive operators, so that $\map B$ is the difference of two CP maps. However, these CP maps are not trace-preserving, so to obtain genuine quantum channels, we analyze the operators $\hat{B}^{\pm}$ in more detail. First, note that similar calculations yield the identities 
\begin{align}
\hat B^+\hat B^+&=\frac{d+1}{2}\hat B^+ \nonumber\\
\hat B^-\hat B^-&=\frac{d-1}{2}\hat B^- \nonumber\\
\hat B^+\hat B^-&=0\;,
\end{align}
which implies $B^{\pm}:=\frac{2}{d\pm1}\hat{B}^{\pm}$ are both orthogonal \textit{projectors}. It then follows that
\begin{equation}
\label{eq:choi-proj}
C(\map B)=\frac{d+1}{2} B^+-\frac{d-1}{2} B^-\;
\end{equation}
is the spectral decomposition of $C[\map B]$ into its positive and negative eigenspaces. Moreover, 
\begin{equation}
\label{eq:partial-traces}
\Tr_{S_{1}S_{2}}[B^\pm]=I_3\;,
\end{equation}
which, by the fact that a linear map $\map E:\Lin(S_{\rm in})\to\Lin(S_{\rm out})$ is trace-preserving if and only if $\Tr_{S_{\rm out}}[C(\map E)]=I_{\rm in}$, guarantees the $B^{\pm}$ are the Choi operators of quantum channels $\map B^{\pm}$.
Therefore, we set $\map B^{\pm}:\Lin(S)\to \Lin(S_1S_2)$ to be the quantum channels given by the inverse of the Choi isomorphism applied to $B^{\pm}$, i.e., 
\begin{align}
\map B^{\pm}(\rho)&:= \Tr_{S_{3}}\big[B^{\pm}(I_{12}\otimes\rho^{T})\big]\nonumber\\
&=\frac{2}{d\pm 1}\Pi^{\pm}(I\otimes \rho)\Pi^{\pm} \quad \forall \rho\in \Lin(S),
\end{align}
where $\rho^{T}$ denotes the transpose of $\rho$ in the standard basis. Thus, $\map B^+$ is precisely the universal optimal quantum cloner~\cite{werner-cloning} and $\map B^-$ is the universal anti-symmetrizer. By a calculation similar to that of the above calculation for $C(
\map B)$, we find
\begin{equation}
C(\map B^{\pm})=\frac{2}{d\pm 1}\hat B^{\pm},
\end{equation}
which together with \eqref{CHDCX77} yields
\begin{equation}
C(\map B)=\frac{d+1}{2}C(\map B^{+})-\frac{d-1}{2}C(\map B^{-}).
\end{equation}
The injectivity of the Choi isomorphism then yields the equality 
\begin{equation}
\label{eq:Bpmdecomp}
\map B=\frac{d+1}{2}\map B^{+}-\frac{d-1}{2}\map B^{-},
\end{equation}
thus concluding the proof of Eq.~\eqref{eq:cloning-decomp}. \qed

We remark that the decomposition in Eq.~\eqref{eq:cloning-decomp}/\eqref{eq:Bpmdecomp} immediately suggests a possible strategy to physically estimate the expectation value of any observable on any state, under the action of the map $\map B$, even though the latter is unphysical. This is because, by linearity, any expectation value $\Tr[\map B(\rho)\, (O_1\otimes O_2)]$ is equal to a linear combination of the expectation values $\Tr[\map B^+(\rho)\, (O_1\otimes O_2)]$ and $\Tr[\map B^-(\rho)\, (O_1\otimes O_2)]$, each of which can be obtained via the action of \textit{physical} quantum channels, namely, the maps $\map B^{\pm}$. This idea was proposed in Ref.~\cite{BDOV13}, which also presented a simple optical implementation for the case $d=2$.

\section{Proof of Theorem~\ref{theo:closest-physical}}
\label{app:closest-physical}

Using the notation from Appendix~\ref{app:spectral-decompproof}, we now compute the diamond norm $\|\map B\|_\diamond$ of the canonical broadcasting map $\map B$. First, we have 
\begin{align}
    \|\map B\|_\diamond&\ge \frac1d\;\big\|C(\map B)\big\|_1\nonumber\\
    &=\frac1d\left(\frac{d+1}{2}\Tr[ B^+]+\frac{d-1}{2}\Tr[ B^-]\right)\nonumber\\
    &=\frac1d\left(\frac{d(d+1)}2+\frac{d(d-1)}2\right)\nonumber\\
    &=d\,.
\end{align}
The first inequality follows from the definition of the diamond norm \eqref{eq:diamond-def} as a maximization over bipartite states, one of which is $\frac{1}{d}\Omega=\frac{1}{d}\sum_{i,j=1}^{d}|i\>\<j|\otimes|i\>\<j|$. The first equality follows from the spectral decomposition \eqref{eq:choi-proj} into the mutually perpendicular positive and negative operators $\frac{d+1}{2}B^{+}$ and $-\frac{d-1}{2}B^{-}$ together with the fact that the trace norm equals the sum of the singular values, which are the absolute values of the eigenvalues in this case. The second equality follows from \eqref{eq:partial-traces}.

On the other hand, Theorem 3 of Ref.~\cite{RTG21} states that given a Hermitian-preserving linear map $\map L$ proportional to a HPTP map (i.e., $\map L^*(I)\propto I$), its diamond norm can be computed as
\begin{equation}
\label{eq:TRG}
\|\map L\|_\diamond=\min\{\lambda_++\lambda_-\}\;,
\end{equation}
where the minimum is taken over all pairs of non-negative real numbers $\lambda_+,\lambda_-\ge 0$ such that there exist two CPTP maps $\map E^+$ and $\map E^-$ satisfying $\map L=\lambda_+\map E^+-\lambda_-\map E^-$. The decomposition \eqref{eq:cloning-decomp}/\eqref{eq:Bpmdecomp} together with \eqref{eq:TRG} then yields
\begin{equation}
    \|\map B\|_\diamond\le \frac{d+1}{2}+\frac{d-1}{2}=d\,.
\end{equation}
Therefore, $\|\map B\|_\diamond=d$.

Now let $\map E$ be an arbitrary CPTP linear map. By invoking the reverse triangle inequality, we have
\begin{equation}
\label{eq:BmEdnorm}
\|\map B-\map E\|_\diamond
\ge \Big| \|\map B\|_\diamond - \|\map E\|_\diamond \Big|
=d-1\,
\end{equation}
where the equality follows from the fact that the diamond norm of a CPTP map equals $1$. Theorem~\ref{theo:closest-physical} is then proved once we show that the above lower bound $\|\map B-\map E\|_\diamond
\ge d-1$ is achieved if and only if $\map E=\map B^+$. Indeed, 
\begin{align}
\|\map B-\map B^+\|_\diamond&=\frac{d-1}{2}\|\map B^+-\map B^-\|_\diamond\nonumber\\
&\le\frac{d-1}{2}\big(\|\map B^+\|_\diamond+\|\map B^-\|_\diamond\big)\nonumber\\
&=d-1
\end{align}
by the triangle inequality and since the diamond norm of a CPTP map is $1$.
This together with \eqref{eq:BmEdnorm} proves $\|\map B-\map B^+\|_{\diamond}=d-1$, so that $\map B^+$ achieves the lower bound given by \eqref{eq:BmEdnorm}. 

Conversely, suppose $\map E$ is a CPTP map such that $\|\map B-\map E\|_{\diamond}=d-1$, and let $\map E^{\pm}$ be the maps given by $\map E^{\pm}(\rho)=\Pi^{\pm}\map E(\rho)\Pi^{\pm}$ for all $\rho$. Momentarily setting
\begin{equation}
Y^{\pm}=C\left(\frac{d\pm 1}{2}\map B^{\pm}\mp\map E^{\pm}\right), 
\end{equation}
we then have
\begingroup
\allowdisplaybreaks
\begin{align}
    d-1&=\|\map B-\map E\|_{\diamond}\nonumber\\
    &\ge \|\map B-\map E^+-\map E^-\|_{\diamond}\nonumber\\
   &\ge\frac1d \big(\| Y^+ - Y^-\|_{1}\big)\nonumber\\
  &=\frac{1}{d}\big(\|Y^{+}\|_{1}+\|Y^-\|_{1}\big)\nonumber\\
   &=\frac{1}{d}\big(\|Y^+\|_{1}+\Tr[Y^-]\big)\nonumber\\
    &\ge \left| \frac{d+1}{2}-\frac{1}{d}\Tr\big[\map E^+(I)\big] \right| +\frac{d-1}{2}+\frac{1}{d}\Tr\big[\map E^-(I)\big]\nonumber\\
    &\ge \frac{d+1}{2}-1+\frac{d-1}{2}+\Tr\left[\map E^-\left(\frac{I}{d}\right)\right] \nonumber\\
    &= d-1+\Tr\left[\map E^-\left(\frac{I}{d}\right)\right]\nonumber\\
    &\ge d-1\;.\label{eq:longinequality}
\end{align}
\endgroup
The first inequality in~\eqref{eq:longinequality} in follows from the contractivity of the diamond norm under CPTP post-processing, i.e., $\|\map L\|_\diamond\ge\|\Phi\circ\map L\|_\diamond$ for all HP $\map L$ and all channels $\Phi$, and taking in particular the pinching post-processing $\Phi(\rho):=\Pi^+\rho\Pi^++\Pi^-\rho\Pi^-$, which satisfies $\Phi\circ(\map B-\map E)=\map B-\map E^{+}-\map E^{-}$. 
The second equality follows from the fact that $Y^+$ and $Y^-$ have mutually orthogonal supports. 
The third equality follows from the fact that $Y^{-}$ is a positive operator so that the trace norm equals the trace. The third inequality follows from the reverse triangle inequality and since $C(\map E^+)$ is positive. 
As for the last two inequalities, we used the fact that both $\map E^+$ and $\map E^-$ are trace-non-increasing completely positive linear maps.
Hence, for any channel $\map E$ saturating the lower bound \eqref{eq:BmEdnorm}, it must be that $\Tr[\map E^-(I/d)]=0$, i.e., $\map E^-=0$, or equivalently, $\map E(\rho)=\map E^+(\rho) \equiv \Pi^+\map E(\rho)\Pi^+$ for all $\rho\in\St(S)$.

Using this fact, it is possible to see that, for any $\map E$ saturating the lower bound, we have
\begin{equation}
    (\map B-\map E)(\rho)=\Pi^+\big(I\otimes \rho-\map E(\rho)\big)\Pi^+-\frac{d-1}{2}\map B^-(\rho)\;,
\end{equation}
for all $\rho$. Hence, the difference $\map B-\map E$ separates into two orthogonal blocks, so that
\begin{equation}
    \|\map B-\map E\|_\diamond
    =\left\|\frac{d+1}{2}\map B^+-\map E\right\|_\diamond+\frac{d-1}{2}\;,
\end{equation}
since $\|\map B^-\|_{\diamond}=1$.
Thus, any channel $\map E$ saturating the bound $\|\map B-\map E\|_{\diamond}=d-1$ must also satisfy
\begin{equation}
    \left\|\frac{d+1}{2}\map B^+-\map E\right\|_\diamond=\frac{d-1}{2}\;.
\end{equation}
Since $\|\map E\|_\diamond=1$, as a consequence of the (conditions for saturation of the) reverse triangle inequality, we conclude that $\map E=\map B^+$.
\qed

\section{Hermitian operator-valued measures} 
\label{app:HOVM}

The definition of an HOVM provided in our Letter in the finite-outcome case can be generalized to arbitrary measurable spaces as follows.  Let $S$ be a quantum system with underlying finite-dimensional Hilbert space $\spc H$ and let $({\sf X},\sigma ({\sf X}))$ be a measurable space, where $\sf X$ denotes the underlying set of the measurable space and $\sigma ({\sf X})$ is a $\sigma$-algebra of measurable subsets of ${\sf X}$. Mimicking the definition of a POVM as in Ref.~\cite{DaLe70} and that of a signed measure as in Ref.~\cite{Rudin87}, a finite {\em Hermitian operator-valued measure} (HOVM) on $({\sf X},\sigma ({\sf X}))$ for $S$ is a function $M : \sigma ({\sf X}) \to  \Herm(S)$ mapping measurable subsets $B\in \sigma ({\sf X})$ into Hermitian operators $M(B)$ satisfying the {\em normalization condition} $M({\sf X})=  I$, the {\em finiteness condition} $\lVert M(B)\rVert<\infty$ for all $B\in\sigma({\sf X})$, and the {\em countable disjoint sum condition} $M\left(\bigcup_{n=1}^{\infty} B_{n}\right)=\sum_{n=1}^{\infty}M(B_{n})$ for any countable family $\{B_{n}\}$ of pairwise disjoint $B_{n}\in\sigma({\sf X})$. Given such an HOVM $M$ and a virtual state $\rho$, the assignment sending $B\in\sigma({\sf X})$ to $\mu_{M}(B):=\Tr\big[\rho M(B)\big]$ defines a finite signed measure $\mu_{M}$ on $({\sf X},\sigma ({\sf X}))$. 

One way to construct an HOVM is by specifying its density with respect to an ordinary probability measure as follows. Let $\mu$ be a probability measure on $({\sf X},\sigma({\sf X}))$ and let $M:  x\in{\sf X}\mapsto  M_x \in \Herm(S)$ be an integrable function in the sense that the function sending $x\in {\sf X}$ to $\<v|M_{x}|w\>$ is $\mu$-integrable for all $v,w\in\spc H$. Then for each $B\in\sigma({\sf X})$, setting $M (B)  :  =  \int_{B} \mu  (\d x)  \,  M_x$ defines an HOVM in the sense above (there is a slight abuse of notation here since $M$ is used as both the integrable function and the induced HOVM). 

The preceding construction is precisely what is used in the first virtual M\&P protocol in this Letter. Namely, the measurable space $({\sf X},\sigma({\sf X}))$ is the Borel space obtained from the topological space of orthogonal rank-one projections in $S$ (which can be thought of as the complex projective space $\mathbb{C}\mathbb{P}^{d-1}$~\cite{BeZy06}). We equip  $({\sf X},\sigma({\sf X}))$ with the probability measure given by the pushforward of the Haar measure on the compact Lie group $\grp U$ of unitary operators on $\spc H$ (the {\em Haar measure} on a compact Lie group is the unique normalized measure that is left invariant under the left action of the group on itself by multiplication). More precisely, given any fixed unit length vector $|v\>\in\spc H$, let $\pi_v$ be the (surjective) measurable function sending a unitary $U\in\grp U$ to the orthogonal rank-one projector $\pi_v(U):=U|v\>\<v|U^{\dag}$. The pushforward measure on $({\sf X},\sigma({\sf X}))$ is then given by $\mu(B)=\mu_{H}(\pi_{v}^{-1}(B))$, where $\mu_{H}$ is the Haar measure on $\grp U$. The probability measure $\mu$ on $({\sf X},\sigma({\sf X}))$ is itself invariant under the adjoint action of $\grp U$ in the sense that $\mu(UBU^{\dag})=\mu(B)$ for all $B\in\sigma({\sf X})$ and $U\in\grp U$ (it is the typical probability measure used to generate Haar random pure states~\cite{BeZy06}). 

Meanwhile, the integrable function $M:{\sf X}\to\Herm(S)$ is given by sending a rank-one projector $\psi$ to the Hermitian operator $M_{\psi}=\frac{d}{2}\big[(d+2)\psi-I\big]$. The function $M$ is indeed integrable because for every $|v\>,|w\>\in\spc H$, the function sending $\psi\in{\sf X}$ to $\frac{d}{2}\big[(d+2)\<v|\psi|w\>-\<v|w\>\big]$ is a bounded and continuous function on a compact space. Therefore, the assignment sending $B\in\sigma({\sf X})$ to $M(B):=\int_{B}\mu(\d \psi)M_{\psi}$ defines the HOVM used in the first M\&P protocol of this Letter, where $\d \psi$ is used as a shorthand for $\mu(\d \psi)$.

\section{Proof of Theorem~\ref{theo:convex}} 
\label{app:theoconvex}

 To prove Theorem~\ref{theo:convex}, it is convenient to use the   Jamio\l kowski  representation~\cite{Ja72}.  For a given linear map $\map L :  \Lin  (S_{\rm in}) \to \Lin  (S_{\rm out})$, the {\em Jamio\l kowski  operator} of $\map L$ is the operator  $J(\map L) \in  \Lin  (S_{\rm in} S_{\rm out})$ defined by 
\begin{align}
 J  (\map L)    & :=      (\map I_{\rm in}  \otimes \map L )   (\tt SWAP_{\rm in, in}  )    \, ,
  \end{align}
where ${\tt SWAP}_{\rm in , in }$ is the swap operator on two identical copies of system $S_{\rm in}$. For the canonical broadcasting map $\map B$, the Jamio\l kowski  operator of $\map B$ is given by 
 \begin{align}
J (\map B)   = \frac 12  \bigg\{      {\tt SWAP}_{12}  \otimes I_3   \, ,     I_1   \otimes {\tt SWAP}_{23}   \bigg\}\,,
 \end{align}
where we have used number subscripts to denote the corresponding tensor factor. 

Now let $\map M$ be the virtual M\&P protocol given by \eqref{MSCPX71}.  Defining $a:  =  d+2$, the Jamio{\l}kowski operator of $\map M$ is given by 
\begin{equation}
\label{JM}
J( \map M)=\frac{d}{8}\big(a^3J_3-a^2J_2+aJ_1-J_0\big),
\end{equation}
where
\begingroup
\allowdisplaybreaks
\begin{align}
J_{3}&=\int \psi \otimes\psi \otimes \psi ~ \d \psi \nonumber\\
J_{2}&=\int ( I \otimes\psi \otimes \psi +  \psi \otimes  I   \otimes \psi+  \psi \otimes\psi \otimes I) ~ \d \psi \nonumber\\
J_{1}&=\int ( \psi \otimes  I \otimes I  + I \otimes \psi \otimes    I+   I \otimes  I \otimes \psi   ) ~ \d\psi \nonumber\\
J_{0}&= I \otimes I \otimes I .
\label{eq:Jiterms}
\end{align}
\endgroup

To compute the various $J_{i}$, we recall the identity 
\begin{equation}
\binom{d+M-1}{M}\int \psi^{\otimes M}~\d\psi=\frac{1}{M!}\sum_{\pi\in S_{M}}U^{(M)}_{\pi}, 
\end{equation}
which comes from combining Eq.\ (3) and the equation before Eq.\ (21) in Ref.~\cite{chiribella2010quantum}. Here, $S_{M}$ is the symmetric/permutation group on the set $\{1,\dots, M\}$ and $U^{(M)}_{\pi}$ is its unitary representation as acting on permuting the tensor factors of $\spc H\otimes\cdots\otimes\spc H$ by that same permutation. 
Setting $\gamma_{ij}:={\tt SWAP}_{ij}$ as the single permutation swapping factors $i$ and $j$, and setting $M$ to be $1,2,3$, we obtain the relations 
\begin{equation}
\label{eq:intHaarpsi}
\int \psi\,\d \psi    =  \frac I d,
\end{equation}
\begin{equation}
\label{eq:intHaarpsipsi}
\int  \psi  \otimes \psi \,\d \psi   =  \frac 1{d(d+1)}  \,   ( I_1\otimes I_2  +  \gamma_{12})  \, , 
\end{equation}  
and 
\begin{equation}
\label{eq:intHaarpsipsipsi}
\int \psi  \otimes \psi  \otimes \psi ~ \d \psi= \frac{1}{d(d+1)(d+2)} (\gamma_{\rm odd}+\gamma_{\rm even}), 
\end{equation}
where
\begin{align}
\gamma_{\rm odd}&=  \gamma_{12} \otimes I_3 + I_1 \otimes \gamma_{23} +    \gamma_{13} \otimes I_2 \nonumber\\
\gamma_{\rm even}&=\left( \gamma_{12} \otimes I_3 \right)  \left( \gamma_{12} \otimes I_3 + I_1 \otimes \gamma_{23}  +    \gamma_{13} \otimes I_2\right)
\end{align}
are the terms coming from the odd and even permutations of $S_{3}$, respectively. Note that $\gamma_{13}\otimes I_{2}$ is short-hand notation for the operator 
\begin{equation}
\label{eq:gamma13I2}
\gamma_{13}\otimes I_{2}:=(\gamma_{12}\otimes I_{3})(I_{1}\otimes\gamma_{23})(\gamma_{12}\otimes I_{3}).
\end{equation}
Inserting Eqs.~\eqref{eq:intHaarpsi} and~\eqref{eq:intHaarpsipsi} into Eq.~\eqref{eq:Jiterms} yields
\begingroup
\allowdisplaybreaks
\begin{align}
J_{2}&=I_{1}\otimes\left(\int\psi\otimes\psi~\d\psi\right)\nonumber\\
&\;\;+(\gamma_{12}\otimes I_{3})\left(I_{1}\otimes\int\psi\otimes\psi~\d\psi\right)(\gamma_{12}\otimes I_{3})
\nonumber\\
&\;\;+\left(\int\psi\otimes\psi~\d\psi\right)\otimes I_{3}\nonumber\\
&=\frac{1}{d(d+1)}\big(3\, I_{1}\otimes I_{2}\otimes I_{3}+\gamma_{\rm odd}\big)
\end{align}
\endgroup
by Eq.~\eqref{eq:gamma13I2}
and
\begin{equation}
J_{1}=\frac{3}{d}\,I_{1}\otimes I_{2}\otimes I_{3}.
\end{equation}
Combining these with Eq.~\eqref{eq:intHaarpsipsipsi} and plugging in the results into Eq.~\eqref{JM} yields 
\begingroup
\allowdisplaybreaks
\begin{align}
&J(\map M) = \frac{d}{8}\bigg(\frac{a^3}{d(d+1)(d+2)}(\gamma_{\rm odd}+\gamma_{\rm even}) \nonumber\\
&\qquad\qquad-\frac{a^2}{d(d+1)}\big(3 I_{123}+\gamma_{\rm odd}\big)+\frac{3a}{d}I_{123}-I_{123}\bigg)\nonumber\\
&=\left[\frac{a^3}{8(d+1)(d+2)}\right](\gamma_{\rm even}-I_{123})\nonumber\\
&\;+\left[\frac{a^3}{8(d+1)(d+2)}-\frac{a^2}{8(d+1)}\right]\gamma_{\rm odd}\nonumber\\
&\;+\left[\frac{a^3}{8(d+1)(d+2)}-\frac{3a^2}{8(d+1)}+\frac{3a}{8}-\frac{d}{8}\right]I_{123}, 
\label{eq:JMstep1}
\end{align}
\endgroup
where $I_{123}:=I_{1}\otimes I_{2}\otimes I_{3}$. 
Now, since Eq.~\eqref{eq:gamma13I2} gives
\begin{equation}
(I_{1}\otimes \gamma_{23})(\gamma_{12}\otimes I_{3})=(\gamma_{12}\otimes I_{3})(\gamma_{13}\otimes I_{2}),
\end{equation}
 we have
\begin{equation}
\{\gamma_{12}\otimes I_{3},I_{1}\otimes\gamma_{23}\}=\gamma_{\rm even}-I_{123}.
\end{equation}
Combining this with the relation $a=d+2$ and simplifying terms in Eq.~\eqref{eq:JMstep1} yields
\begin{align}
 J(\map M) &=\frac{(d+2)^2}{8(d+1)}\{\gamma_{12}\otimes I_{3},I_{1}\otimes\gamma_{23}\}-\left[\frac{1}{4(d+1)}\right]I_{123}\nonumber\\
 &=\left[\frac{(d+2)^2}{4(d+1)}\right]\frac{\{\gamma_{12}\otimes I_{3},I_{1}\otimes\gamma_{23}\}}{2}\nonumber\\
 &\;\;+\left[1-\frac{(d+2)^2}{4(d+1)}\right]\frac{I_1\otimes I_2\otimes I_3}{d^2}\,.
\end{align}
 
Recalling that $J(\map B)  =    \{   \gamma_{12} \otimes I_3    \, ,     I_1 \otimes \gamma_{23}  \}/2$ and noting that the virtual M\&P protocal corresponding to the completely depolarizing channel $\map M'$ given by \eqref{DSCPX87} has Jamio{\l}kowski operator $J(\map M')  =   (I_1\otimes I_2\otimes I_3)/d^2$, we then obtain the identity
\begin{align}
J(\map M)   =     \frac{(d+2)^2}{4(d+1)} \,   J( \map B )   +      \bigg[  1- \frac{(d+2)^2}{4(d+1)}   \bigg]    J(\map M') \,.
\end{align}  
Solving for $J(\map B)$ then yields  
\begin{align}
\nonumber J(\map B) &  =     \frac{J(\map M)   +   \bigg[   \frac{(d+2)^2}{4(d+1)}  -1  \bigg]   \,  J(\map M') }{  \frac{(d+2)^2}{4(d+1)}  } \\
\nonumber &  =  p  J(\map M)   +  (1-p)  \,  J(\map M')  \, \\
  &  = J \big(  p \,\map M +  (1-p)\,   \map M' \big)
\, , \quad p:  =  \frac{4(d+1)}{(d+2)^2}\, ,
\end{align}
where linearity of the Jamio{\l}kowski isomorphism was used. 
The injectivity of the Jamio{\l}kowski isomorphism then yields the equality $\map B  =  p\, \map M +  (1-p)\, \map M'$, thus concluding the proof. \qed

\section{The broadcasting condition for state over time functions} 
\label{app:broadcastingcondition}

We recall that the broadcasting condition for a state over time function $\star$ is the assumption that for every $(\map E,\rho)$, there exists a broadcasting map $\map B$ such that
\begin{equation}
\map E\star \rho=(\map I\otimes \map E)(\map B(\rho).
\end{equation}
In this appendix, we show that the broadcasting condition for a quantum state over time can be derived as a consequence of consistency with probabilistic mixtures of states and consistency with post-processing via arbitrary quantum channels. We also give a more operational interpretation in terms of the Heisenberg picture.

First, a state over time function $\star$ is said to be \emph{convex-linear} whenever 
\begin{equation}\label{eq:convexity}\map E\star \left(\sum_ip_i\rho_i\right)  = \sum_{i}  \,  p_i \,  \map  E \star \rho_i, 
\end{equation}  
for all convex combinations $\sum_i \, p_i\,  \rho_i$, where  $(\rho_i)_i$ is a finite collection of states and $(p_i)_i$ is a probability distribution. 

Second, a state over time function $\star$ is said to be \emph{consistent with post-processing} whenever 
\begin{equation}\label{eq:postprocessing}
(\map F\circ \map E)\star\rho=(\map I\otimes\map F)(\map E\star\rho)
\end{equation}  
for all quantum channels $\map E$, $\map F$, and for all states $\rho$. 

\begin{prop}
\label{prop:broadcastingSOT}
If a state over time function $\star$ is convex linear and consistent with post-processing, then it satisfies the broadcasting condition.
\end{prop}

\medskip

\noindent
{\bf Proof} Setting $\map F=\map E$ and $\map E=\map I$ in the post-processing condition \eqref{eq:postprocessing} yields $\map E\star\rho=(\map I\otimes\map E)(\map I\star\rho)$. The statement then follows once we show that the map $\map B$ given by $\map B(\rho)=\map I\star \rho$ is a broadcasting map. Indeed, $\map B$ is convex-linear by \eqref{eq:convexity}, and it also satisfies $\Tr_{S_1}\circ \hspace{1mm}\map B=\Tr_{S_2}\circ \hspace{1mm} \map B=\map I$ by the marginal condition \eqref{MCXD67} for states over time. Hence, $\map B$ is a broadcasting map.
\qed 

We now provide a more operational alternative to the post-processing axiom for states over time. Namely, a state over time function $\star$ is said to be \emph{consistent with the Heisenberg picture} whenever 
\begin{align}
\label{eq:heisenberg}
\Tr_{S_2} &\Big[\big(I  \otimes  \map F^{*}  (P_i)   \big)   (  \map E \star \rho)   \Big]  \nonumber \\
&= \Tr_{S_2}  \Big[(I  \otimes  P_i     )  \big((\map F \circ \map E ) \star \rho\big) \Big] \,  
\end{align}
for all $i$, 
every state $\rho$, every POVM $(P_i)_i$, and every pair of composable channels $\map E$ followed by $\map F$. Here, $\map F^{*}$ is the Hilbert--Schmidt adjoint of $\map F$. 

Condition~\eqref{eq:heisenberg} can be justified on physical terms by considering the following measurement scheme on the state over time $\map E\star\rho$. For this, let the first and second factors associated with a state over time correspond to Alice and Bob, respectively. First, Bob performs a quantum channel $\map F$ on the virtual state $\map E\star\rho$, and then measures the output of this channel with a POVM $(P_i)$. Mathematically, the overall measurement of the channel $\map F$ and POVM  $(P_i)$ is described by the POVM $(\map F^{*} (P_i))_i$. It then becomes natural to expect that the induced state on Alice's system obtained upon Bob performing such a measurement $(\map F^{*} (P_i))_i$ on the second system of the state over time $\map E\star \rho$ should be equivalent to Bob performing the measurement $(P_i)$ on the second system of the state over time  $(\map F\circ\map E) \star \rho$. This condition is mathematically captured by~\eqref{eq:heisenberg}. 

\begin{prop}
\label{prop:broadcastingSOTHeisenberg}
If a state over time function $\star$ is convex linear and consistent with the Heisenberg picture, then it satisfies the broadcasting condition.
\end{prop}

\medskip

\noindent
{\bf Proof} By the arguments of Proposition~\ref{prop:broadcastingSOT}, it suffices to prove the equivalence between \eqref{eq:postprocessing} and \eqref{eq:heisenberg}. 

Assume \eqref{eq:postprocessing} holds. Then
\begin{align}
\Tr_{S_2}&\Big[(I\otimes \map F^*(P_i))(\map E\star \rho)\Big]\nonumber\\
&=\Tr_{S_2}\Big[\big((\map I\otimes \map F^*)(I^{\dag}\otimes P_i^{\dag})\big)^{\dag}(\map E\star \rho)\Big] \nonumber \\
&=\Tr_{S_2}\Big[(I\otimes P_i)^{\dag}\big((\map I\otimes \map F)(\map E\star \rho)\big)\Big] \nonumber \\
&=\Tr_{S_2}\Big[(I\otimes P_i)\big((\map F\circ \map E)\star \rho\big)\Big], 
\end{align}
where the first equality follows from the fact that $\map I\otimes\map F^*$ is Hermitian-preserving, the second equality follows from the definition of the Hilbert--Schmidt adjoint, and the last equality follows from~\eqref{eq:postprocessing}. 
This proves that \eqref{eq:postprocessing} implies \eqref{eq:heisenberg}.

Conversely, suppose \eqref{eq:heisenberg} holds. Let $A_{1}\in\Lin(S_1)$ and $A_{2}\in\Lin(S_2)$ be arbitrary positive operators. Then
\begin{align}
&\Tr\bigg[(A_{1}\otimes A_{2})\big((\map I\otimes\map F)(\map E\star\rho)\big)\bigg] \nonumber\\
&=\Tr_{S_{1}}\Big[(A_{1}\otimes I)\Tr_{S_{2}}\Big[(I\otimes A_{2})\big((\map I\otimes \map F)(\map E\star\rho)\big)\Big]\nonumber\\
&=\Tr_{S_{1}}\bigg[(A_{1}\otimes I)\Tr_{S_{2}}\Big[\big(I\otimes\map F^*(A_{2})\big)(\map E\star\rho)\Big]\bigg]\nonumber\\
&=\Tr_{S_1}\bigg[(A_{1}\otimes I)\Tr_{S_2}\Big[(I\otimes A_{2})\big((\map F\circ\map E)\star\rho\big)\Big]\bigg]\nonumber\\
&=\Tr\Big[(A_{1}\otimes A_{2})\big((\map F\circ\map E)\star\rho\big)\Big],
\label{eq:TrA1A2POVM}
\end{align}
where the first equality holds by properties of the partial trace, the second equality follows from the fact that $A_{2}$ is Hermitian and the definition of the Hilbert--Schmidt adjoint, the third equality follows from \eqref{eq:heisenberg} and linearity (by rescaling $A_{2}$ by its operator norm, one has $\frac{A_{2}}{\| A_{2}\|}\le I$ so that it can always be viewed as part of a two-element POVM), and the fourth equality follows by the properties of the partial trace. Since~\eqref{eq:TrA1A2POVM} holds for every positive $A_{1},A_{2}$, and because every operator is a linear combination of four positive elements, it follows that 
\begin{align}
\Tr\bigg[(A_{1}&\otimes A_{2})\big((\map I\otimes\map F)(\map E\star\rho)\big)\bigg] \nonumber\\
&=\Tr\Big[(A_{1}\otimes A_{2})\big((\map F\circ\map E)\star\rho\big)\Big]
\end{align}
for arbitrary $A_{1}\in\Lin(S_1)$ and $A_{2}\in\Lin(S_2)$. By the faithfulness of the trace, this is equivalent to \eqref{eq:postprocessing}, thus concluding the proof. \qed

 \bibliography{references}

\end{document}